\documentclass[preprint,12pt,authoryear]{elsarticle}

\usepackage{amssymb}

\journal{Journal of Non-Newtonian Fluid Mechanics }

\begin{document}

\begin{frontmatter}



\title{Rheological behaviour of suspensions of bubbles in yield stress fluids}


\author{Lucie Duclou\'e}
\cortext[cor]{corresponding author}
\ead{lucie.ducloue@ifsttar.fr}
\author{Olivier Pitois}
\author{Julie Goyon}
\author{Xavier Chateau}
\author{Guillaume Ovarlez}
\address{Laboratoire Navier (UMR CNRS 8205), Universit\'e Paris-Est, 77420 Champs-sur-Marne, France}

\begin{abstract}
The rheological properties of suspensions of bubbles in yield stress fluids are investigated through experiments on model systems made of monodisperse bubbles dispersed in concentrated emulsions. Thanks to this highly tunable system, the bubble size and the rheological properties of the suspending yield stress fluid are varied over a wide range. We show that the macroscopic response under shear of the suspensions depends on the gas volume fraction and the bubble stiffness in the suspending fluid. This relative stiffness can be quantified through capillary numbers comparing the capillary pressure to stress scales associated with the rheological properties of the suspending fluid. We demonstrate that those capillary numbers govern the decrease of the elastic and loss moduli, the absence of variation of the yield stress and the increase of the consistency with the gas volume fraction, for the investigated range of capillary numbers. Micro-mechanical estimates are consistent with the experimental data and provide insight on the experimental results.
\end{abstract}


\end{frontmatter}


\section{Introduction}
Yield stress fluids are widely used in the industry where their versatile character, from solid under a critical stress to liquid above that threshold, has many applications~\citep{coussot2005rheometry}. Examples include creams and gels in the cosmetic industry, and also mud or fresh building materials like plaster or concrete slurries. During processing of those materials, air bubbles are often present in the fluid, either because they get entrapped during mixing or as the result of deliberate addition to confer innovative properties to the final product. This is for instance the case in dairy products~\citep{vanAken2001333} or in the building industry in which aerated materials are designed to be lighter and better insulating. Processing of these aerated yield stress fluids requires to understand and monitor their behaviour under shear flows.

\vspace{0.5cm}
Understanding the response of a sheared bubble suspension in a non-Newtonian fluid is complex: as the suspending fluid itself is non-Newtonian, the behaviour of the suspension is expected to be non-Newtonian too, and the contribution of additional non-linear phenomena due to the presence of the bubbles may be difficult to quantify. Some useful understanding of the physical mechanisms at stake can be collected from previous results on related cases of simpler suspensions. The simplest type of suspension is a dispersion of solid particles in a Newtonian fluid. The relative viscosity of such suspensions is an increasing function of the solid volume fraction which is well described by a Krieger-Dougherty law~\citep{wildemuth1984viscosity}. From a microscopic point of view, all the shear deformation undergone by the suspension occurs in the fluid between the solid grains. As a consequence, the effective local shear rate in the fluid has to be greater than the macroscopic shear rate applied to the suspension, leading to increased dissipation.
\newline
Suspensions of bubbles in Newtonian fluids have been studied by \cite{rust2002bubble} and \cite{llewellin2002rheology}: their experiments showed that the relative viscosity of the bubbly liquid in a steady shear flow increases with the gas volume fraction at low shear rate (with a lesser growth than the relative viscosity of particle suspensions) and decreases at high shear rate. Observation of the bubbles in the flow (also quantified in~\cite{rust2002effects}) evidenced the importance of bubble deformation in the contribution of the bubbles to the overall viscosity: bubbles in their experimental set-up are spherical at low shear rate and elongated in the flow at high shear rate. The distortion of flow lines around non-deformable bubbles leads to increased local shear rates in the suspending fluid compared to the macroscopic shear rate applied to the suspension. However, the absence of friction at the bubble surface lessens the total dissipation in the bubble suspension compared to the particle suspension. At high shear rate, the elongation of inviscid bubbles in the flow accommodates part of the shear deformation and decreases the total dissipation. This transition from stiff to soft bubbles with increasing shear rate is the result of a competition between two physical effects: the viscous stress in the fluid tends to stretch the bubbles in the flow whereas the capillary stress minimizes the bubbles' surface by favouring a spherical shape. To quantify this competition, the authors introduce a capillary number that can be defined as ``viscous'' and is the ratio of the viscous stress to the capillary stress: $Ca_{visc}=\frac{\eta \dot{\gamma}}{\sigma/R}$ where $\eta$ is the viscosity of the suspending fluid, $\dot{\gamma}$ is the applied shear rate, $\sigma$ is the surface tension between the gas and the liquid and $R$ is the bubble radius.
\newline
The case of suspensions in non-Newtonian fluids is more complicated as the local shear rate between the particles, and consequently the apparent viscosity of the interstitial fluid, is not known. Numerous experiments have been performed on filled polymer melts, which are dispersions of rigid particles in visco-elastic fluids and have large industrial applications~\citep{mewis2012colloidal}. The results obtained by~\cite{poslinski1988rheological} on suspensions of glass spheres in a polymer melt shed light on two important effects of particle addition in a non-Newtonian fluid. In the absence of fillers, the suspending fluid considered by the authors is Newtonian at low shear rate, and then shear-thinning for higher shear rates. When particles are added to the fluid, the viscosity of the suspension is increased for all shear rates, and the Newtonian plateau gets shorter and shorter as the solid volume fraction increases. The onset of shear-thinning in the suspension for lower shear rates is due to shear amplification in the fluid between the particles, in which the effective local shear rate can be high enough to get off the Newtonian plateau even though at the macroscopic shear rate applied to the suspension the fluid alone would still be Newtonian. The overall response of the suspension is the coupling of the fluid rheology to the flow lines perturbation caused by the inclusions. 
Suspensions of particles in yield stress fluids have been studied by \cite{mahaut2008yield} who characterized the elastic and plastic response of suspensions of hard spheres in a Herschel-Bulkley fluid. Below the yield stress, the elastic modulus of the suspensions grows with the solid volume fraction and follows a Krieger-Dougherty law, as can be expected from the viscosity of suspensions in Newtonian fluids: both measurements characterize the linear response of each suspension. The yield stress increases with the solid volume fraction too, and its growth, of smaller magnitude than the one of the linear properties of the suspensions, is well predicted by micro-mechanical estimates~\citep{chateau2008homogenization}. For shear thinning yield stress fluids, the lesser growth of the yield stress compared to the linear properties of the suspension can also be understood as a manifestation of shear amplification in the fluid between the grains. The local shear rate in the suspending fluid is higher than the macroscopic shear rate applied to the suspension and increases with the solid volume fraction, leading to decreasing apparent (secant) viscosity of the interstitial fluid. The overall response results from the interplay of flow lines perturbation and apparent fluidification of the suspending fluid.


\vspace{0.5cm}
Bubbly yield stress fluids have been the subject of fewer studies. Besides stability studies~\citep{goyon2008spatial, C1SM06537H}, their elastic properties have been studied in detail in~\cite{duclouecoupling2014} and a first description of the rheology of mixtures of foams and pastes has been given in \cite{kogan2013mixtures}. However, more work is needed to investigate a broader range of rheological parameters for the yield stress fluid and to describe the flow properties of those suspensions. We anticipate from the results on bubble suspensions in Newtonian fluids that the rheology of an aerated yield stress fluid will also be the result of the interplay of the suspending fluid rheology and capillary forces acting on the bubble surface. Yield stress fluids behave as visco-elastic solids below their yield stress and visco-plastic fluids above that threshold. We are thus interested in the visco-elastic properties, yield stress and flow curve of suspensions of bubbles in those fluids. In this aim, we perform an experimental study of the overall rheological properties of model suspensions of bubbles in tunable yield stress fluids. We limit to gas volume fractions up to 50\% so that we do not consider foams of yield stress fluids, in which the bubbles are deformed by geometrical constraints. 

\vspace{0.5cm}
In section~\ref{section:Mats}, we present the materials used for the study, and the rheometrical procedures. In section~\ref{section:Complex modulus}, we discuss the complex shear modulus of a soft aerated solid. In section~\ref{section:Yield stress}, we review our results for the plasticity threshold of bubbly yield stress fluids. Section~\ref{section:Flow consistency} is dedicated to the flow characterisation of our suspensions.

\section{Materials and methods}
\label{section:Mats}
To perform this experimental study, we prepare model suspensions of monodisperse bubbles in simple yield stress fluids. For most systems, the suspensions are obtained by mixing a simple yield stress fluid with a separately produced monodisperse foam.
\subsection{Model yield stress fluids}
The simple yield stress fluids that we choose to perform the study are concentrated oil in water emulsions. By changing the chemical composition of the two phases and the oil volume fraction, we obtain various suspending emulsions with elastic moduli ranging from 100 to 1000Pa and yield stresses between 10 and 40Pa. Unless otherwise indicated, the radius of the droplets is around 1 to 2$\mathrm{\mu m}$ (the polydispersity, computed as in~\cite{mabille2000rheological}, is around 20\%). At the considered gas volume fractions, this small droplet size should ensure that there is scale separation between the drops and the bubbles, and consequently validate the use of the suspending emulsion as a continuous medium embedding the bubbles~\citep{goyon2008spatial}. The variety of the suspending emulsions used for the study is illustrated in table~\ref{tab:recap_systs}, which gives their composition. 
\begin{center}
\begin{table}
{\renewcommand{\arraystretch}{1.5}
\renewcommand{\tabcolsep}{0.2cm}
\begin{scriptsize}
\begin{tabular}{|l|c|c|c|}
\hline
& \textbf{oil - vol. fraction} & \textbf{continuous phase} & \textbf{$\mathbf{\sigma}$ (mN.$\mathbf{m^{-1}}$})\\
\hline
emulsion (1) & dodecane - 73\% & SDS 2.7\% w. in water & 36 $\pm$ 1\\
\hline
emulsion (2a) & silicon (V20) - 75\% & Forafac\textregistered ($\mathrm{{DuPont}^{TM}}$) 4\% w. in water & 15.5 $\pm$ 0.1\\
\hline
emulsion (2b) & silicon (V20) - 73\% & Forafac\textregistered ($\mathrm{{DuPont}^{TM}}$) 4\% w. in water & 15.5 $\pm$ 0.1\\
\hline
emulsion (3) & silicon (V350) - 79\% & TTAB 3\% w. in water/glycerol 50/50 w/w & 35.5 $\pm$ 0.1\\
\hline
emulsion (4a) & silicon (V350) - 70\% & TTAB 3\% w. in water/glycerol 36/64 w/w & 35 $\pm$ 1\\
\hline
emulsion (4b) & silicon (V350) - 70\% & TTAB 3\% w. in water/glycerol 36/64 w/w & 35 $\pm$ 1\\
\hline
\end{tabular}
\end{scriptsize}}
\caption{Synthetic description of all the concentrated emulsions used as model yield stress fluids to prepare bubble suspensions: nature and volume fraction of the oil dispersed phase, composition of the aqueous continuous phase (including the surfactant) and surface tension between the air and the continuous phase.\label{tab:recap_systs}}
\end{table}
\end{center}
\subsection{Suspensions preparation}
\begin{figure}[h!]
\centering
\includegraphics[scale=0.2]{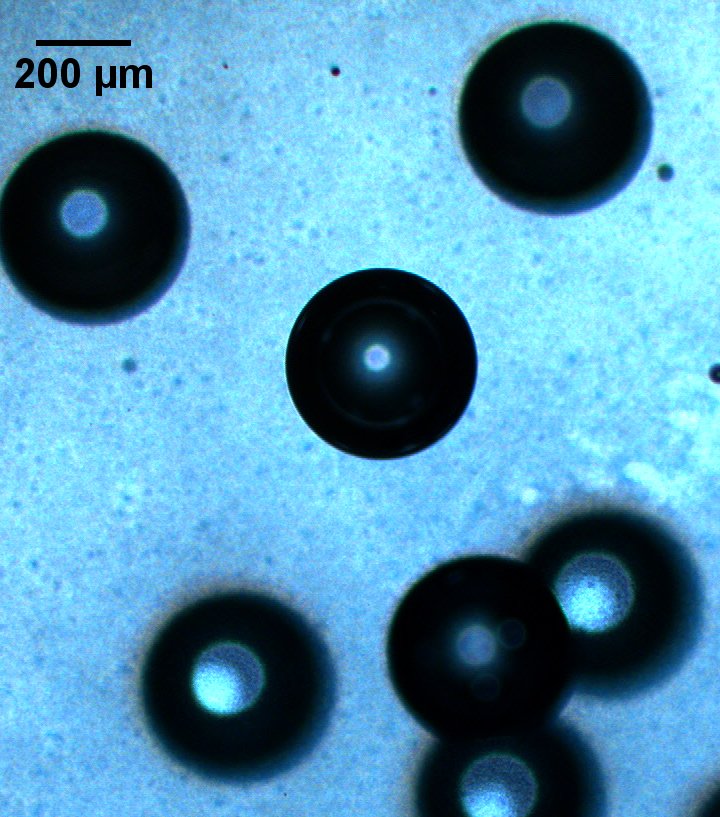}
\caption{Microphotograph of a bubble suspension in suspending emulsion 3. The bubble radius is 200$\mu$m. The granulated background is emulsion 3, which is transparent.\label{fig:photo}}
\end{figure}

Most suspensions are prepared by gently mixing the suspending emulsion with a separately produced monodisperse foam. The foams are obtained by blowing nitrogen plus a small amount of perfluorohexane ($\mathrm{C_6F_{14}}$) through a porous glass frit or through needles: we are able to produce nearly monodisperse foams with average bubble radii $R$ ranging from 40$\mathrm{\mu m}$ to 800$\mathrm{\mu m}$. Coarsening is strongly reduced by the presence  of $\mathrm{C_6F_{14}}$~\citep{gandolfo1997interbubble}, meaning that the bubble size is stable during measurements. The continuous phase of the foam is the same as the one in the emulsion, ensuring that the dispersion of bubbles in the emulsion during mixing is easy and does not induce any additional chemical effect in the suspensions. The mixing with the foam adds in a small amount of continuous phase to the emulsion, which lowers its oil volume fraction and thus alters its rheological behaviour. For a series of suspensions at different gas volume fractions in a given emulsion, we fix the oil volume fraction of the suspending emulsion by adding as necessary a complement of pure continuous phase in the system (the same protocol was used in~\cite{PhysRevLett.104.128301, kogan2013mixtures} for instance). An example of a dispersion of bubbles in an emulsion is shown in figure~\ref{fig:photo}.
The prepared bubble suspension is then poured in the rheometer geometry to perform measurements. Because of the random nature of the mixing and pouring process, we assume that the suspension is isotropic when it is set in place in the rheometer.
\subsection{Rheometrical procedures}
\begin{figure}[h!]
\centering
\includegraphics[scale=0.25]{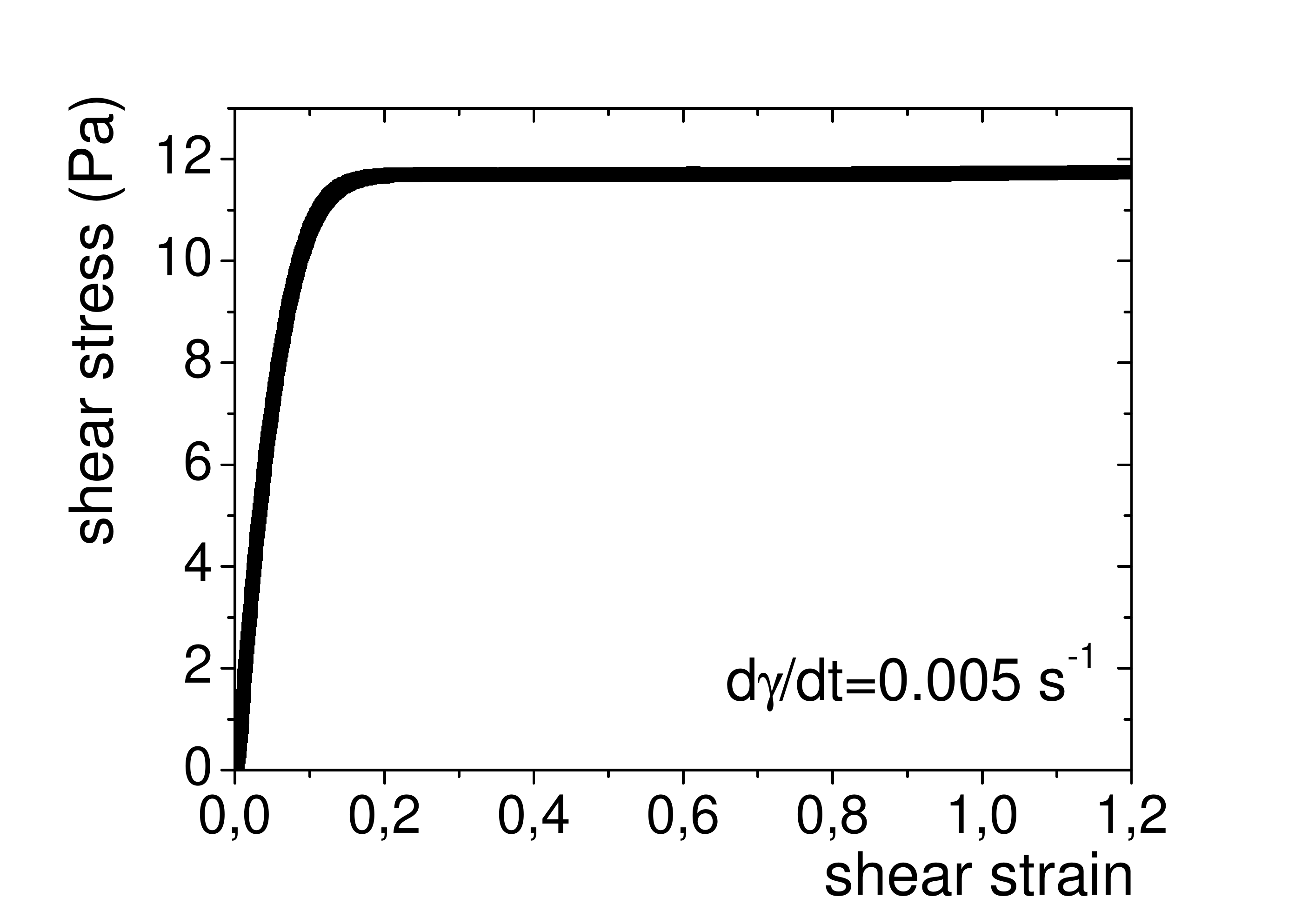}\includegraphics[scale=0.25]{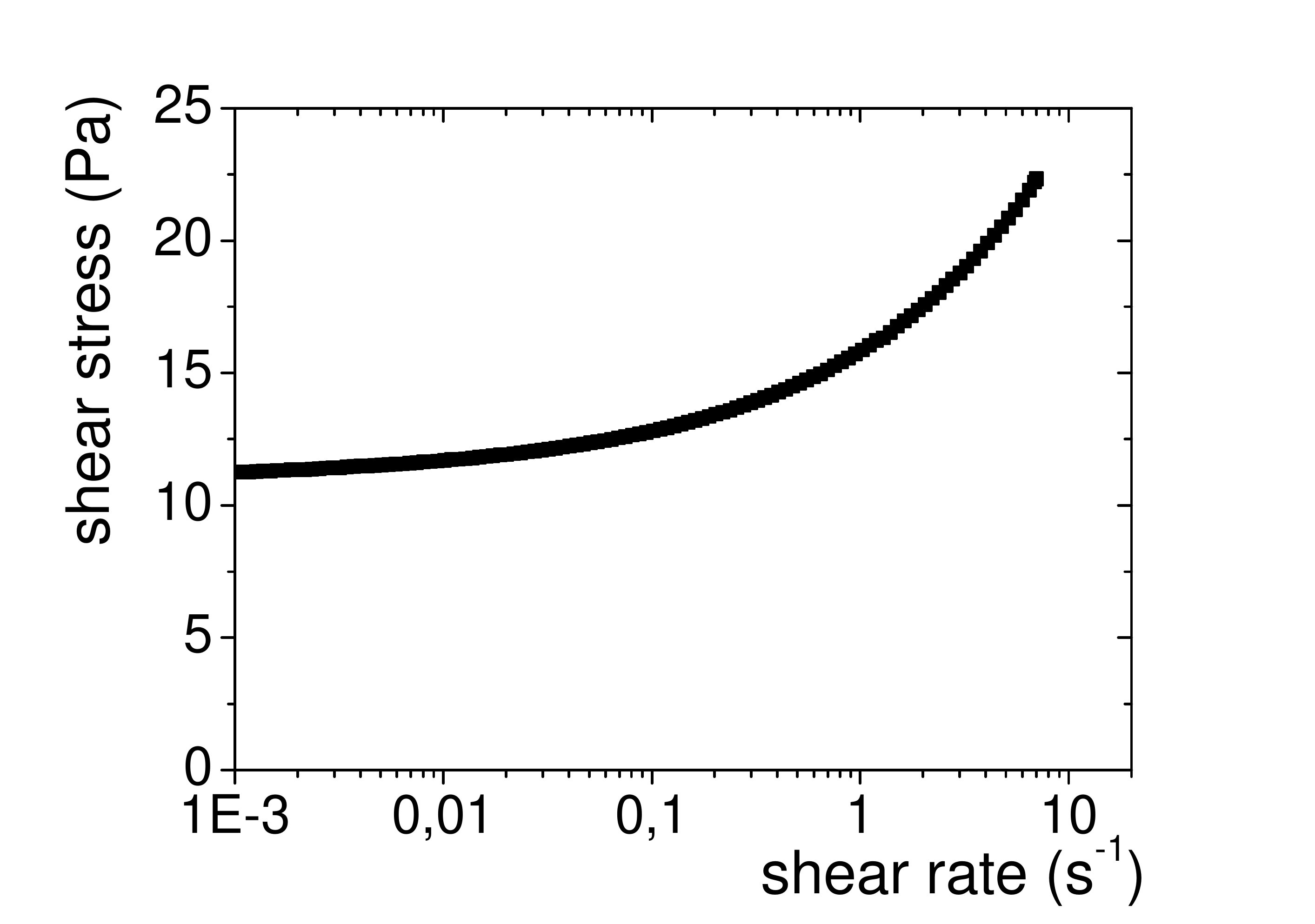}
\caption{Yield stress measurement \textit{(left)} and flow curve \textit{(right)} of suspending emulsion 2a. The material is elasto-plastic below the yield stress and visco-plastic beyond.\label{fig:seuil}}
\end{figure}

The rheometrical measurements are performed on a stress-controlled rheometer (either Bohlin C-VOR 200 or Malvern Kinexus Ultra). The geometry used to perform the measurement does not affect the result providing the sheared thickness of suspension is larger than several bubble diameters. Depending on the bubble size, different geometries were used to ensure that this condition was satisfied, while minimizing the required volume of material to fill in the geometry. For $R_{b}\le 50\mathrm{\mu m}$, the material is sheared between parallel plates (radius $R$=25mm, gap $h$=2.5mm). The planes are serrated to prevent slippage of the suspension~\citep{coussot2005rheometry}. Suspensions containing bigger bubbles require a larger thickness of sheared material and are studied in roughened Couette-like devices : for $50\mathrm{\mu m}< R_{b}< 800\mathrm{\mu m}$, we use a vane in cup (exceptionally a serrated bob in cup) geometry (inner radius $R_i$=12.5mm, outer radius $R_o$=18mm), and for $R_{b}\ge 800\mathrm{\mu m}$, we use vane in cup geometries (either $R_i$=12.5mm and $R_o$=25mm or $R_i$=22.5mm and $R_o$=45mm). 
\newline
The rheometrical procedure is described below. Typical values of the parameters are given all along the description.
The shear modulus $G'$ of the suspensions is measured immediately after setting the material in the rheometer, by imposing small amplitude oscillations at a frequency of typically 1Hz. The oscillatory stress is chosen to be well below the yield stress of the suspensions, so that the oscillations are performed in the linear elastic regime of each suspension. At this frequency, the loss modulus of the systems is negligible.
\newline
After the elastic modulus, the static yield stress $\tau_y$ of the suspensions is measured by initiating flow from rest at a small and constant imposed shear rate $\dot{\gamma}$, typically $0.005{\mathrm{s}}^{-1}$. The low shear rate ensures that the contribution of viscous effects to the torque is negligible. The curve obtained during the measurement of the static yield stress of suspending emulsion 2a is shown as an example in figure~\ref{fig:seuil}. The suspending emulsion is elasto-plastic: the stress increases first linearly with the strain, until it reaches a plateau at yielding. 
\newline
Once the plateau stress is attained, the suspensions are set to rest at zero stress for a few dozens of seconds, so that the elastic deformation stored in the material is relaxed~\cite{cloitre2000rheological}. After that, a shear rate ramp is applied over around 2 minutes, typically from $10^{-3}$ to 10${\mathrm{s}}^{-1}$, which, with our suspensions, is enough to ensure that the viscous contribution to the total torque overcomes the one of the yield stress. The flow curve of suspending emulsion 2a is shown in figure~\ref{fig:seuil}. The flow curve of all suspending emulsions is well fitted to a Herschel-Bulkley model $\tau(\dot{\gamma})=\tau_y+k{\dot{\gamma}}^{n}$, where $k$ is the consistency and $n\sim0.5$ is the plastic index.
\newline
The rheological characteristics of the suspending emulsions used for the study are detailed in table~\ref{tab:recap_systs2}, along with the radius of the bubbles that were added to each suspending emulsion to prepare the suspensions. The parameters $k$ and $n$ of the flow curve of the suspending emulsions are not discussed here, and are reserved for section~\ref{section:Flow consistency}. For a given set of suspending emulsion and bubble size, several suspensions were produced to vary the gas volume fraction in the range 0\%-50\%. 
\begin{center}
\begin{table}
{\renewcommand{\arraystretch}{1.5}
\renewcommand{\tabcolsep}{0.2cm}
\begin{small}
\begin{tabular}{|c|c|c|c|c|}
\hline
& \multicolumn{2}{|c|}{\textbf{Fluid rheology}} & \multicolumn{2}{|c|}{\textbf{Capillarity}}\\ 
\multicolumn{1}{|l|}{\textbf{Emulsion}} & $G'(0)$ (Pa) & $\tau_y(0)$ (Pa) & $R$ ($\mu$m) & $\sigma/(2R)$ (Pa) \\
\hline
&  &  & 50 $\pm$ 10 & 360 $\pm$ 82\\
1 & 285 $\pm$ 20 & 10 $\pm$ 0.4 & 143 $\pm$ 17 & 120 $\pm$ 18\\
 &  &  & 800 $\pm$ 40 & 23 $\pm$ 1.7\\
\hline
2a & 230 $\pm$ 20 & 12 $\pm$ 0.3 & 41 $\pm$ 5 & 189 $\pm$ 54\\
\hline
2b & 163 $\pm$ 10 & 7.2 $\pm$ 0.5 & 129 $\pm$ 10 & 60 $\pm$ 11\\
\hline
3 & 650 $\pm$ 50 & 40 $\pm$ 2 & 1000 $\pm$ 100 & 18 $\pm$ 2.2\\
\hline
4a & 650 $\pm$ 50 & 20.5 $\pm$ 0.5 & 50 $\pm$ 10 & 350 $\pm$ 80\\
\hline
4b & 799 $\pm$ 40 & 20.5 $\pm$ 0.5 & 150 $\pm$ 10 & 117 $\pm$ 11\\
\hline
\end{tabular}
\end{small}}
\caption{Relevant stress scales for the study: rheological characteristics of the suspending emulsions, and capillary stress scale in the bubbles. The flow curve description is discussed below in section~\ref{section:Flow consistency}.\label{tab:recap_systs2}}
\end{table}
\end{center}

\section{Complex modulus}
\label{section:Complex modulus}
\subsection{Elastic modulus}
\label{ssection:Elastic modulus}
\begin{figure}[h!]
\includegraphics[scale=0.24]{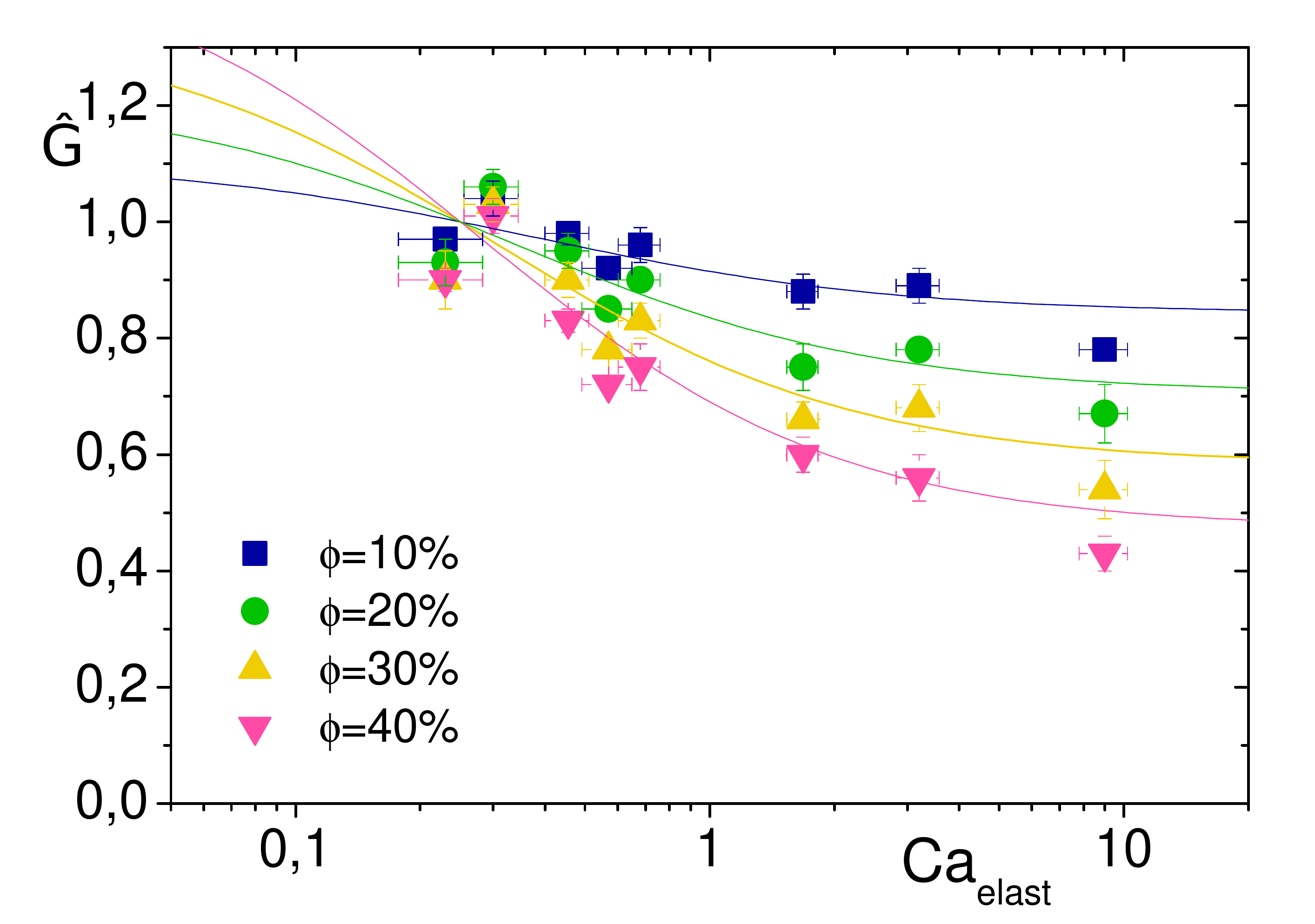}
\includegraphics[scale=0.24]{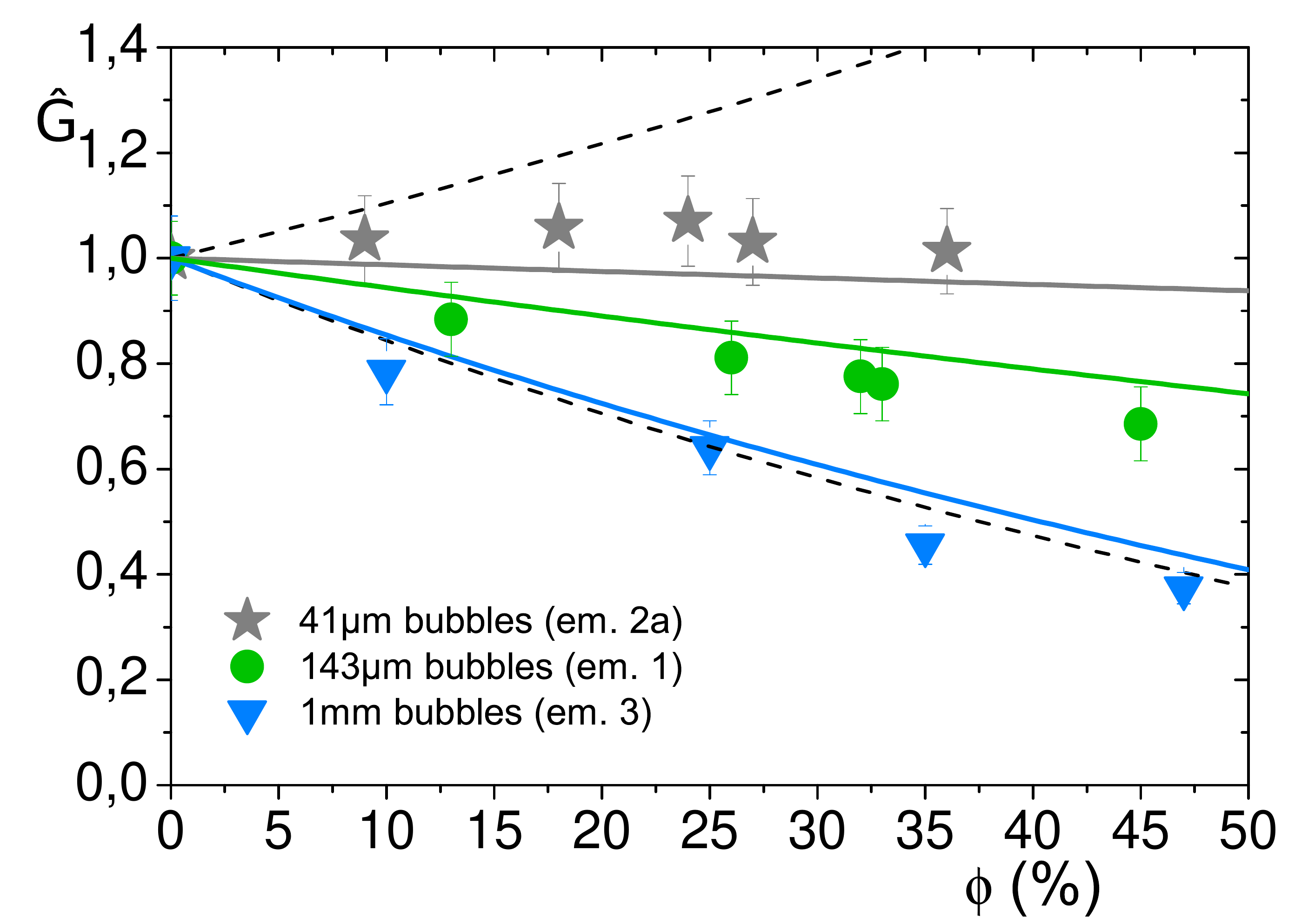}
\caption{\textit{(left)} Dimensionless elastic modulus $\hat{G}$ of the suspensions as a function of the elastic capillary number $Ca_{elast}$ for four values of $\phi$. Symbols are experimental data points, full lines are micro-mechanical computations \textit{(figure from~\cite{duclouecoupling2014})}. \textit{(right)} Dimensionless elastic modulus $\hat{G}$ of 3 bubble dispersions in suspending emulsions as a function of the gas volume fraction $\phi$. Symbols are experimental data points, full lines are micro-mechanical computations; the capillary numbers are, from top to bottom: 0.3, 0.6, 9. From 0.25 to 9, $\hat{G}$ turns from being roughly constant with $\phi$ to decreasing as fast as for surface tension-free pores. Dashed lines are the computed limits from equation~\ref{eq:homog}, for suspensions of infinitely rigid ($Ca_{elast}=0$, top) and freely deformable ($Ca_{elast}\to \infty$, bottom) bubbles \textit{(data replotted from~\cite{duclouecoupling2014})}.\label{fig:elast}}
\end{figure}

\paragraph{Experimental results}
For applied stresses well below the yield stress, the suspending emulsion behaves as a soft visco-elastic solid. The study of the linear elastic properties of suspensions of bubbles in elastic solids has been the subject of previous work~\citep{duclouecoupling2014}. Because of surface tension forces at the interface between the gas and the suspending emulsion, bubbles resist deformation and behave as equivalent soft elastic inclusions in the unyielded suspending emulsion. The overall elasticity of the suspensions depends on the ratio of the fluid's elastic modulus to the bubble's equivalent elasticity in the suspending medium. This competition is quantitatively governed by a capillary number defined as the ratio of the suspending medium elastic modulus to the capillary stress scale in the bubbles, given by the Laplace pressure: 
\begin{equation}
Ca_{elast}=\frac{G'(0)}{2\sigma/R}
\end{equation}
For a suspension of bubbles with known radius in a given suspending emulsion, $Ca_{elast}$ is entirely determined and the elastic modulus $G'(\phi)$ of the suspension depends solely on the gas volume fraction $\phi$. In the range of $Ca_{elast}$ experimentally accessible with our set-up ($0.2 \leq Ca_{elast} \leq9$) $G'$ decreases with $\phi$ and this decrease is all the more significant as $Ca_{elast}$ is large. Micro-mechanical estimates taking into account the physical parameters $\phi$ and $Ca_{elast}$ of the systems predict dimensionless elastic moduli $\hat{G}=G'(\phi)/G'(0)$ for semi-dilute suspensions~\citep{thyoverall, palierne1990linear}:
	\begin{equation}
	\hat{G}_{homog}(\phi, Ca)=1-\frac{\phi(4Ca-1)}{1+\frac{12}{5}Ca-\frac{2}{5}\phi(1-4Ca)}
	\label{eq:homog}
	\end{equation}
in the Mori-Tanaka scheme. This estimate allows to predict values for $\hat{G}$ in a broader range of $Ca$ than is experimentally accessible with our systems. The computed $\hat{G}$ undergoes a transition for $Ca=0.25$: below this value, $\hat{G}$ increases with $\phi$, and it turns into a decreasing function of $\phi$ for $Ca>0.25$. In this latter range of capillary numbers, the micro-mechanical estimates are in good agreement with the experimental measurements. Experimental results from \cite{duclouecoupling2014} as well as micro-mechanical estimates for $\hat{G}$ as a function of $Ca_{elast}$ \textit{(left)} and $\phi$ \textit{(right)} are presented in figure~\ref{fig:elast}. 
\subsection{Viscous modulus}
\begin{figure}[h!]
\includegraphics[scale=0.25]{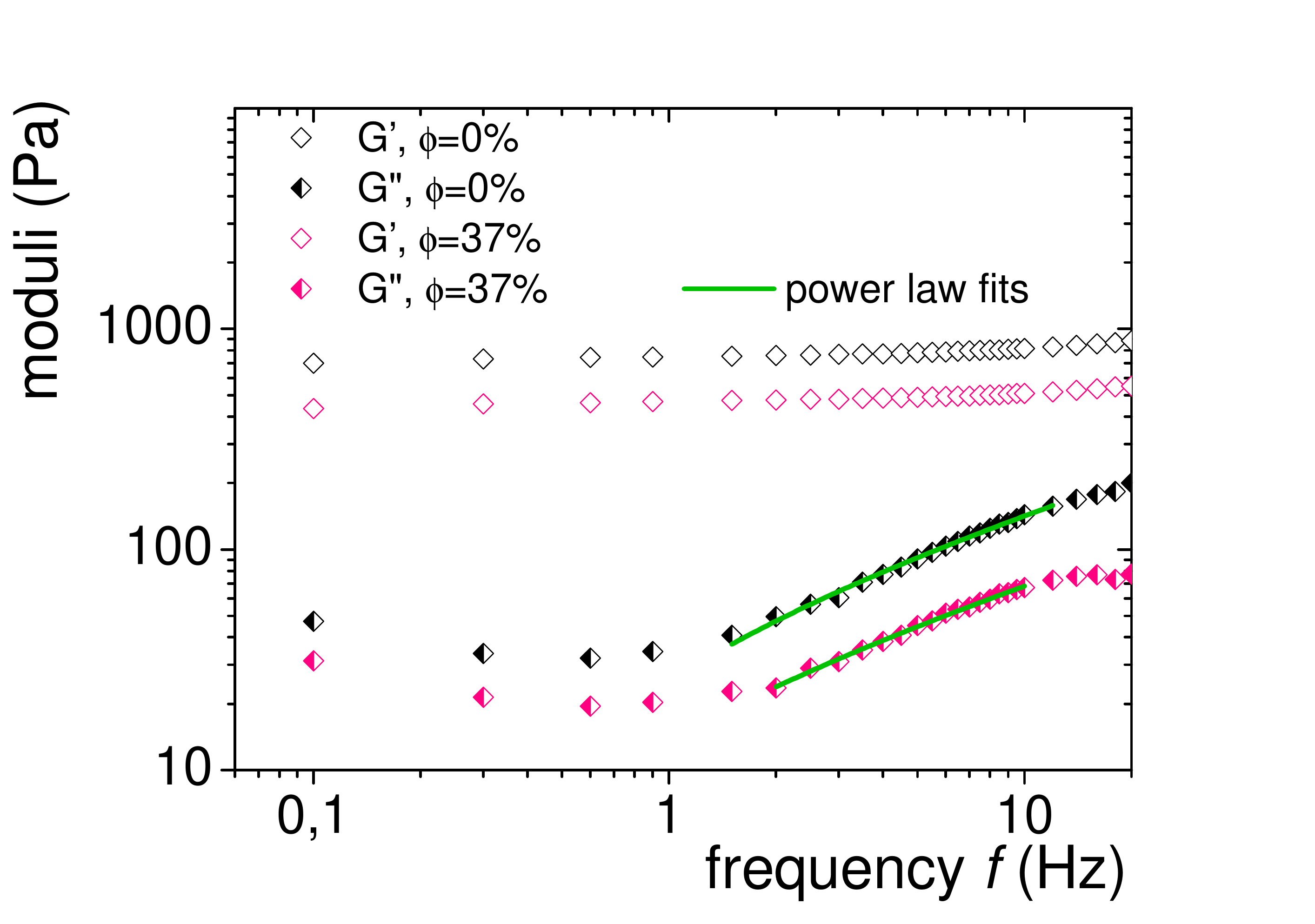}
\includegraphics[scale=0.25]{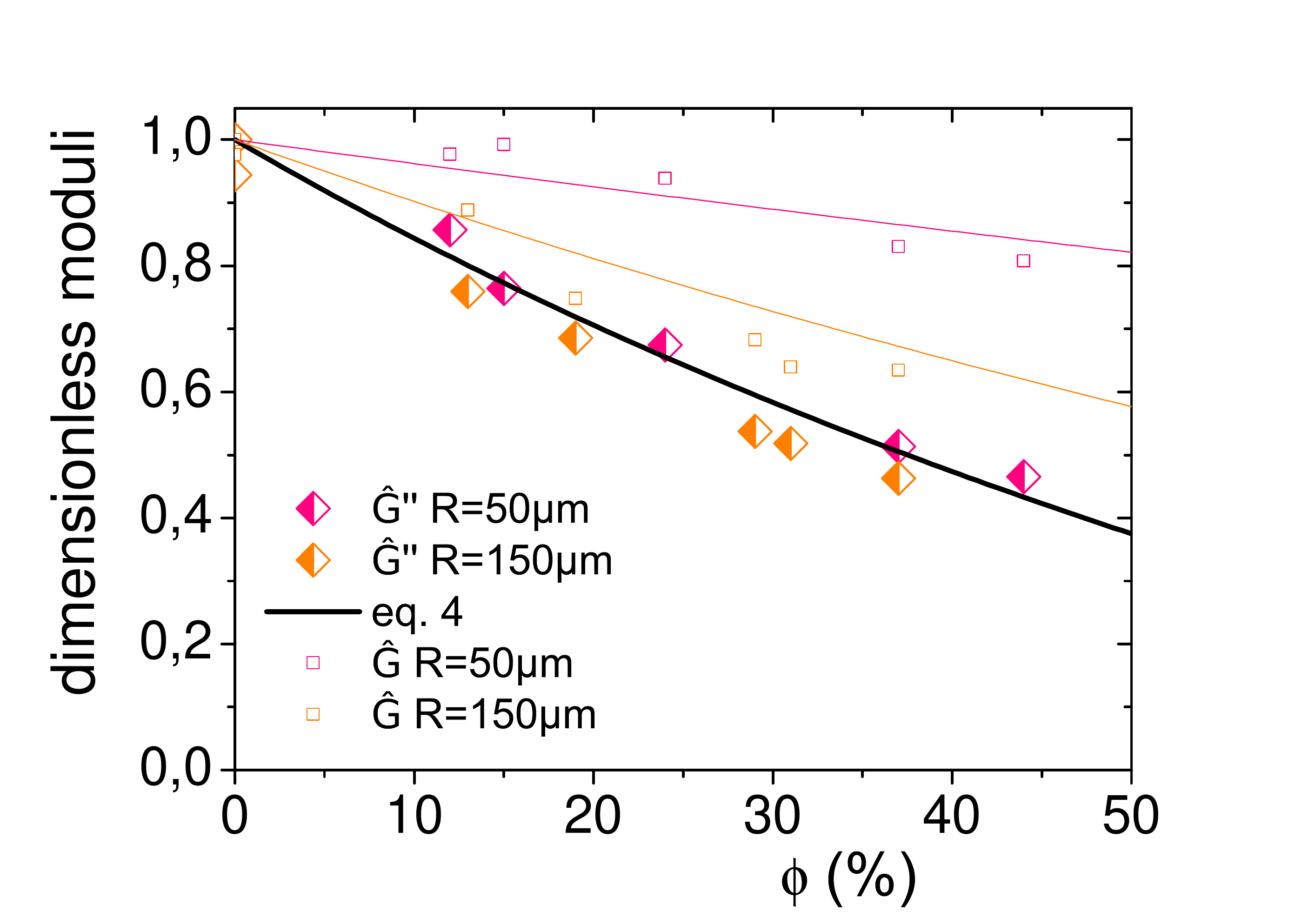}
\caption{\textit{(left)} Frequency dependence of elastic and viscous moduli of suspending emulsion 4b (black symbols) and a suspension of 150$\mu$m bubbles in that emulsion. The green lines are the power law fits to the $G''(f)$. \textit{(right)} Dimensionless elastic and viscous moduli as a function of the gas volume fraction for two suspensions of bubbles in emulsion 4: 50$\mu$m in emulsion 4a and 150$\mu$m in emulsion 4b. The full colored lines are the computed dimensionless elastic moduli for those systems. The thick black line is given by equation~\ref{eq:Ca_inf}.\label{fig:g_loss_adims}}
\end{figure}

The viscous modulus of the systems, defined as the imaginary part of the complex shear modulus, is negligible during the oscillatory measurement performed at 1Hz. To study its evolution with the gas volume fraction, we design a system with high viscous effects by using an aqueous phase with a high glycerol weight content (64\%) (emulsion 4). Because of the high glycerol weight content in the aqueous phase, this emulsion is unstable below 25$^{\circ}$C. The measurements are all performed at 25$^{\circ}$C, but the temperature was poorly controlled during preparation and storage of this emulsion, which may explain that the two batches of this emulsion we made have slightly different moduli. We accurately measure the viscous component of the complex modulus by performing oscillations at very small deformation over a frequency sweep, in the range 0.1Hz $\leq f \leq$ 50Hz. For 1Hz $\leq f \leq$ 20Hz, the viscous modulus $G''(0)$ of the suspending emulsion stands out against the noise in the oscillations and scales as the square root of the frequency: $G''(0)\sim a(0) f^{0.5}$. Above this frequency, we did not manage to calibrate the inertia of our geometry with enough precision to get accurate moduli. This frequency dependence of the loss modulus is known for concentrated emulsions~\citep{mason1995optical}, as well as aqueous foams~\citep{cohen1998viscoelastic}, which are structurally very similar. The power law scaling remains the same for the suspensions of bubbles in the suspending emulsion: $G''(\phi)\sim a(\phi) f^{0.5}$ in the same range of frequency. $G'(f)$ and $G''(f)$ for suspending emulsion 4b and a suspension of bubbles in this emulsion are plotted in figure~\ref{fig:g_loss_adims}. We quantify the evolution of the viscous modulus with the gas volume fraction by computing $\hat{G''}(\phi)=a(\phi)/a(0)$.
\newline
$\hat{G''}(\phi)$ is plotted in figure~\ref{fig:g_loss_adims} for suspensions of bubbles with two different radii in two batches of suspending emulsion 4. $\hat{G}(\phi)$ for the same suspensions is also re-plotted for comparison. We observe that $\hat{G''}$ is a decreasing function of $\phi$. It can be noticed that this decrease is much larger than observed for the elastic modulus of the same suspensions. Remarkably, and contrary to the observations on $\hat{G}$, although $R$ is three times larger in one of the suspensions, the values for $\hat{G''}(\phi)$ are similar for both series of suspensions.

\subsection{Discussion}
\begin{figure}[h!]
\centering
\includegraphics[scale=0.25]{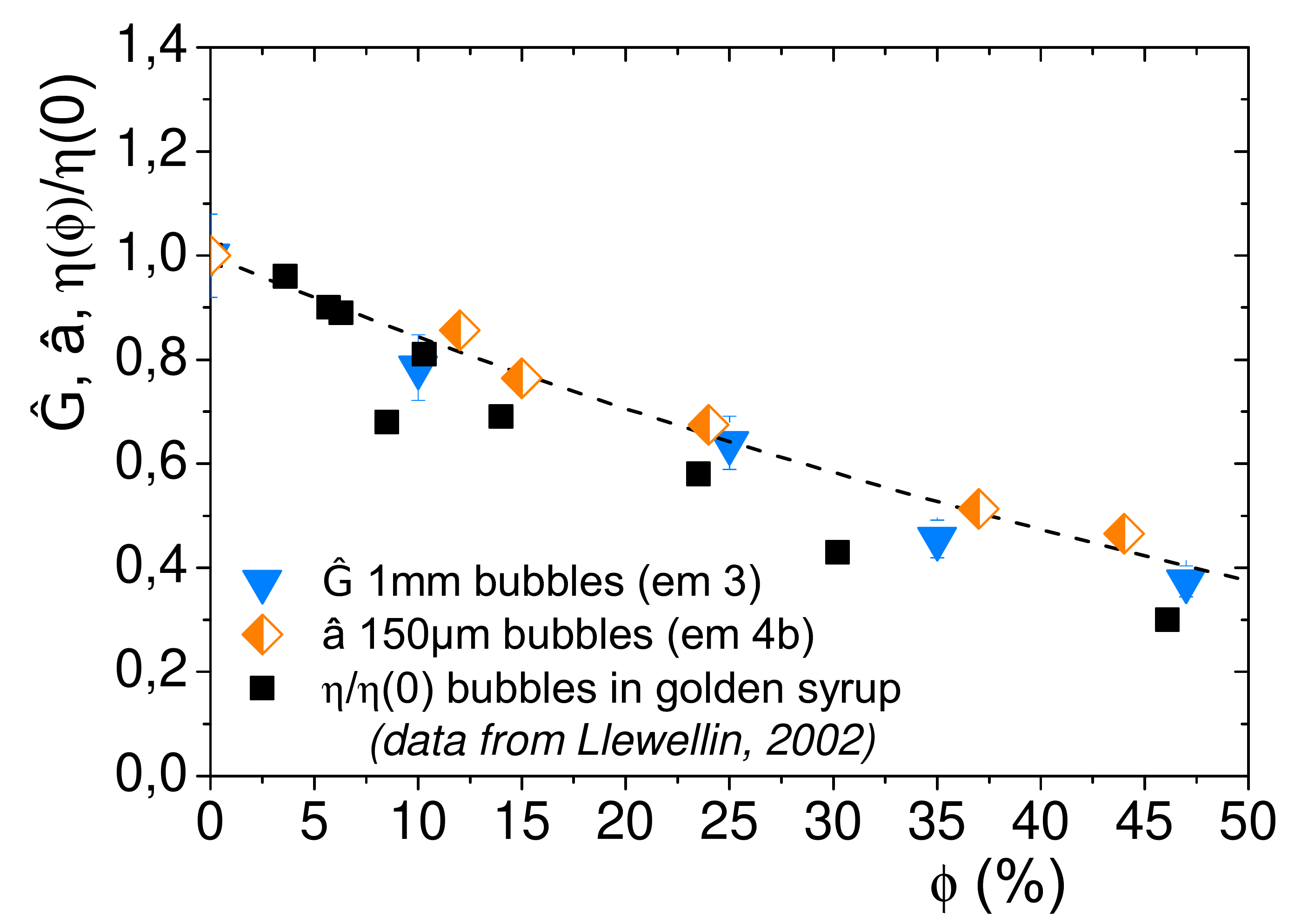}
\caption{Dimensionless viscosity (bubbles in a Newtonian liquid, data replotted from~\cite{llewellin2002rheology}), dimensionless reduced dissipation $\hat{G''}$ and dimensionless elastic modulus (both, bubbles in a yield stress fluid) as a function of $\phi$. The dashed line is, as before, the result of equation~\ref{eq:homog} in the limit case of $Ca \to \infty$.\label{fig:comp_llewellin}}
\end{figure}

\paragraph{Loss capillary number}
To understand that result, we would like to compare $G''(0)$, which is a scale of viscous dissipation in the suspending emulsion, to a scale of viscous dissipation arising from the bubbles. In the suspending emulsion, the dissipated power per unit volume scales as $G''(0){\gamma_o}^{2} f$, where $\gamma_o$ is the oscillating strain amplitude. In the bubbles, the dissipation mainly comes from the change in the area of the bubbles under shear. As the oscillating deformation is very small ($\gamma_o\sim 10^{-4}$), there is no macroscopic surfactant flow on the surface of the bubbles and so the change of a bubble relative area is governed by the macroscopic applied strain $\gamma_o$ and scales as $\gamma_o$~\citep{palierne1990linear}. The surface viscosity at the bubble interface has two contributions: a shear viscosity and a dilatational viscosity. For most surfactant solutions, the contribution of the dilatational viscosity greatly overcomes the one of the shear viscosity~\citep{cohen2013flow}. Surface rheology measurements~\citep{kao1992measurement, biance2009topological} indicate that the surface dilatational viscosity of mobile surfactant solutions (employed in our systems) is of order $\eta_d\sim 10^{-5}$ to $10^{-4}\mathrm{N.s.m}^{-1}$. The dissipated power in a bubble per unit volume is proportional to the bubble specific surface area and so finally it scales as $f \eta_d {\gamma_o}^{2} f /R$. The ratio of the power dissipated per unit volume in the suspending emulsion compared to the one in the bubbles reads 
\begin{equation}
Ca_{loss}=\frac{G''(0)}{f \eta_d / R}
\end{equation}
which defines a loss capillary number. The value of $G''$ for $f \sim 8$Hz, which is located in the power-law scaling part of the experimental curve, is around 95Pa in emulsion 4a and 125Pa in emulsion 4b. For the suspensions of 50$\mu$m bubbles in emulsion 4a, $Ca_{loss}$ is around 6 to 60 depending on the exact value of $\eta_d$ and it is around 23 to 230 for the 150$\mu$m bubbles in emulsion 4b. The high value of $Ca_{loss}$ emphasizes that the dissipation is much higher in the suspending emulsion, which can be qualitatively understood by noticing that dissipation in the emulsion mostly arises from the same mechanisms as in the bubbles, from interfacial solicitation at the droplet surface. As the droplets are much smaller than the bubbles, the droplets' specific surface area is much larger than the bubbles' one and the dissipation per unit volume is higher in the emulsion. This argument is only qualitative, because contrary to the bubbles, the droplets are compressed and dissipation can also occur in the films of continuous phase separating the droplets. Given that the decrease of $\hat{G''}(\phi)$ that we observe in the experiments does not depend on the bubble radius, and thus not on the bubble specific surface area, we can guess that bubble dissipation is likely to be negligible, which in terms of loss capillary numbers means that $Ca_{loss}$ is very high. As the bubbles' contribution to the overall dissipation is very small, we can use the results established above for the elastic modulus (equation~\ref{eq:homog}) in the limit of infinite capillary number, by noticing that both moduli are the linear response of the suspensions and should be described by the same equations providing we use the suitable capillary number. This function 
	\begin{equation}
	\hat{G}_{homog}(\phi, Ca\to\infty)=\frac{1-\phi}{1+(2/3)\phi}
	\label{eq:Ca_inf}
	\end{equation}
is plotted in figure~\ref{fig:g_loss_adims} (thick black line) and is in good agreement with the measured $\hat{G''}$. This function happens to be equal to the Mori-Tanaka bound, which is a classical poromechanics result for the elastic modulus of an elastic solid containing pores with no surface tension~\citep{dormieux2006microporomechanics}.
\paragraph{Special case $Ca \to \infty$}
The moduli for small amplitude oscillations are the linear response of our suspensions of bubbles in yield stress fluids, which we would like to compare to the linear response of suspensions of bubbles in Newtonian fluids, that is: their relative viscosity. The viscosity measurements made by~\cite{rust2002effects} are not suitable for comparison with our experiments because they are performed in a steady shear flow, which means that deformable bubbles are elongated in the flow (whereas to the first order in deformation they remain spherical in our systems). However,~\cite{llewellin2002rheology} have performed oscillatory measurements on the same suspensions of bubbles in a viscous syrup to determine their relative viscosity. Their measurements are made over a frequency sweep at constant shear stress (and consequently constant $\gamma_o f$ with $\gamma_o$ the amplitude of the oscillating strain and $f$ the frequency). At low frequency, $\gamma_o\sim 2$, and the bubbles undergo large deformation, which cannot compare to our experiments. At high frequency, though, the strain is small and the bubbles remain spherical in the liquid. The dimensionless viscosity of their suspensions on the high frequency plateau thus correspond to a similar situation as our moduli measurement at 1Hz. They find a rather fast decrease in this plateau viscosity with $\phi$. Both their experimental results, $\hat{G}$ for our suspensions of very soft bubbles, and $\hat{G''}(\phi)$ for the suspension in emulsion 4b are plotted for comparison in figure~\ref{fig:comp_llewellin}. All three data sets are very close. In all cases, the bubbles's contribution to the measured quantity is very small compared to that of the surrounding medium. This translates into negligible elastic contribution compared to the yield stress fluid, and negligible viscous dissipation compared to the Newtonian fluid and the yield stress fluid. Although the nature of the suspensions is different, the similarity in the equations for the linear response of the systems and the naught contribution of the bubbles yields to similar evolution with $\phi$. For all systems, this evolution is well described by the micro-mechanical estimate of equation~\ref{eq:Ca_inf}.
\section{Yield stress}
\label{section:Yield stress}
\subsection{Experimental results}
\begin{figure}[h!]
\centering
\includegraphics[scale=0.3]{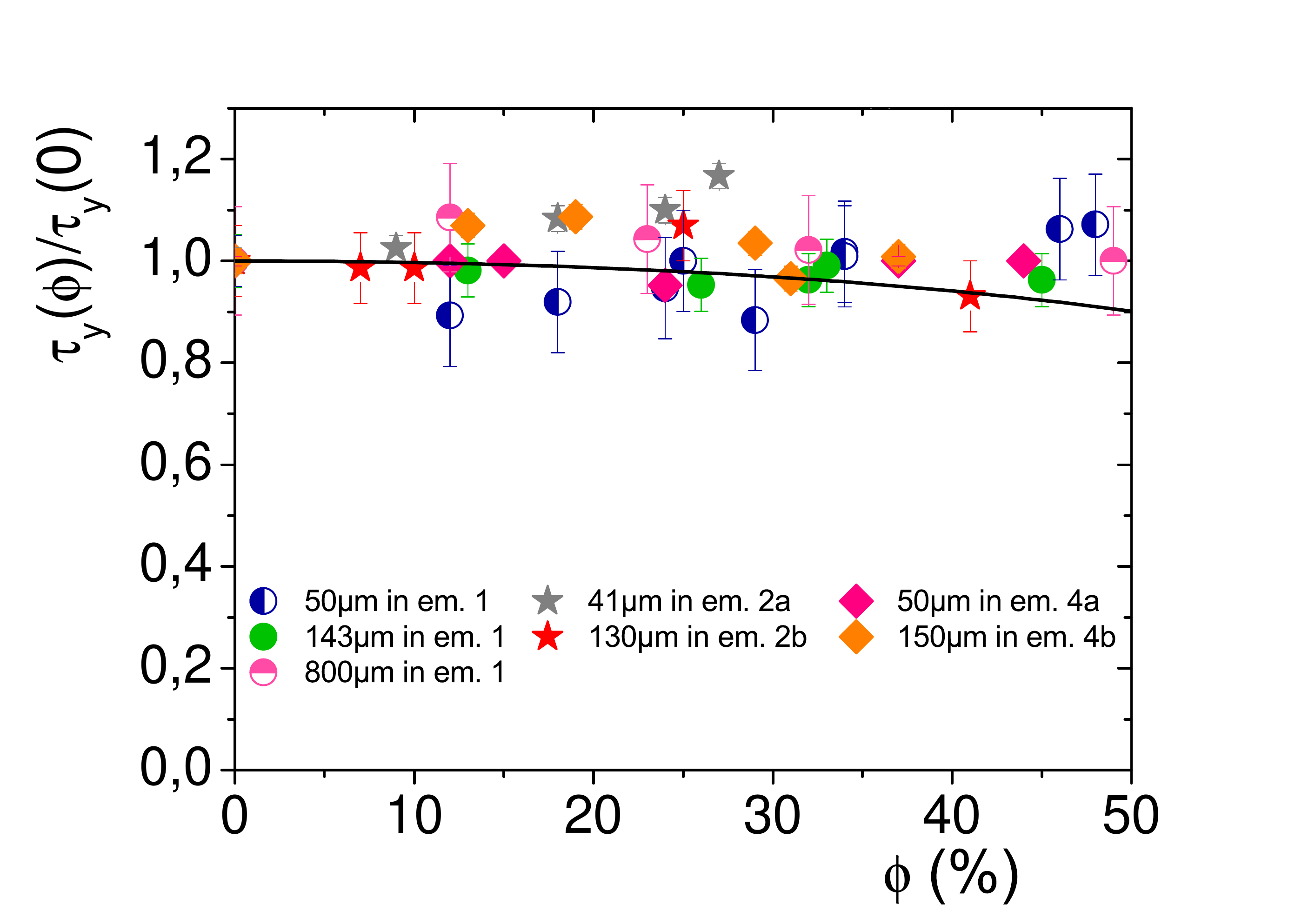}
\caption{Dimensionless yield stress as a function of the gas volume fraction, for seven suspensions detailed in the figure caption: symbols refer to the emulsion composition and colors differentiate the bubble sizes. The line is a micro-mechanical estimate (equation~\ref{eq:tau_ca_0}).\label{fig:seuils_adims}}
\end{figure}

The plastic properties of the suspensions exhibit a very different behaviour from that of the moduli in solid regime. We discuss below the evolution of the plateau yield stress with the gas volume fraction. The shape of the whole stress-strain curve during the yield stress measurement is discussed in Appendix. We measured the yield stress of seven bubble suspensions with bubble radius $R$ ranging from (41$\pm$5)$\mu$m to (800$\pm$40)$\mu$m and suspending emulsions of yield stress $\tau_y(0)$ between 7Pa and 20Pa. Surprisingly, for all these systems, we find that for gas volume fractions $\phi$ up to 50\% the plateau yield stress is not affected by the presence of the bubbles. This remains true all over the radius range mentioned above. The dimensionless yield stress $\tau_y(\phi)/\tau_y(0)$ as a function of $\phi$ is plotted for all these suspensions in figure~\ref{fig:seuils_adims}. Given the experimental precision, $\tau_y(\phi)/\tau_y(0)\sim 1$, meaning that the yield stress of a bubble suspension is comparable to the one of the suspending emulsion for the range of parameters investigated. Neither the emulsion's yield stress, the bubble radius nor the surface tension seem to play a role, although the range of physical and rheometrical parameters explored is rather extensive (see tables~\ref{tab:recap_systs} and~\ref{tab:recap_systs2} for details).
\subsection{Discussion}
\begin{figure}[h!]
\centering
\includegraphics[scale=0.3]{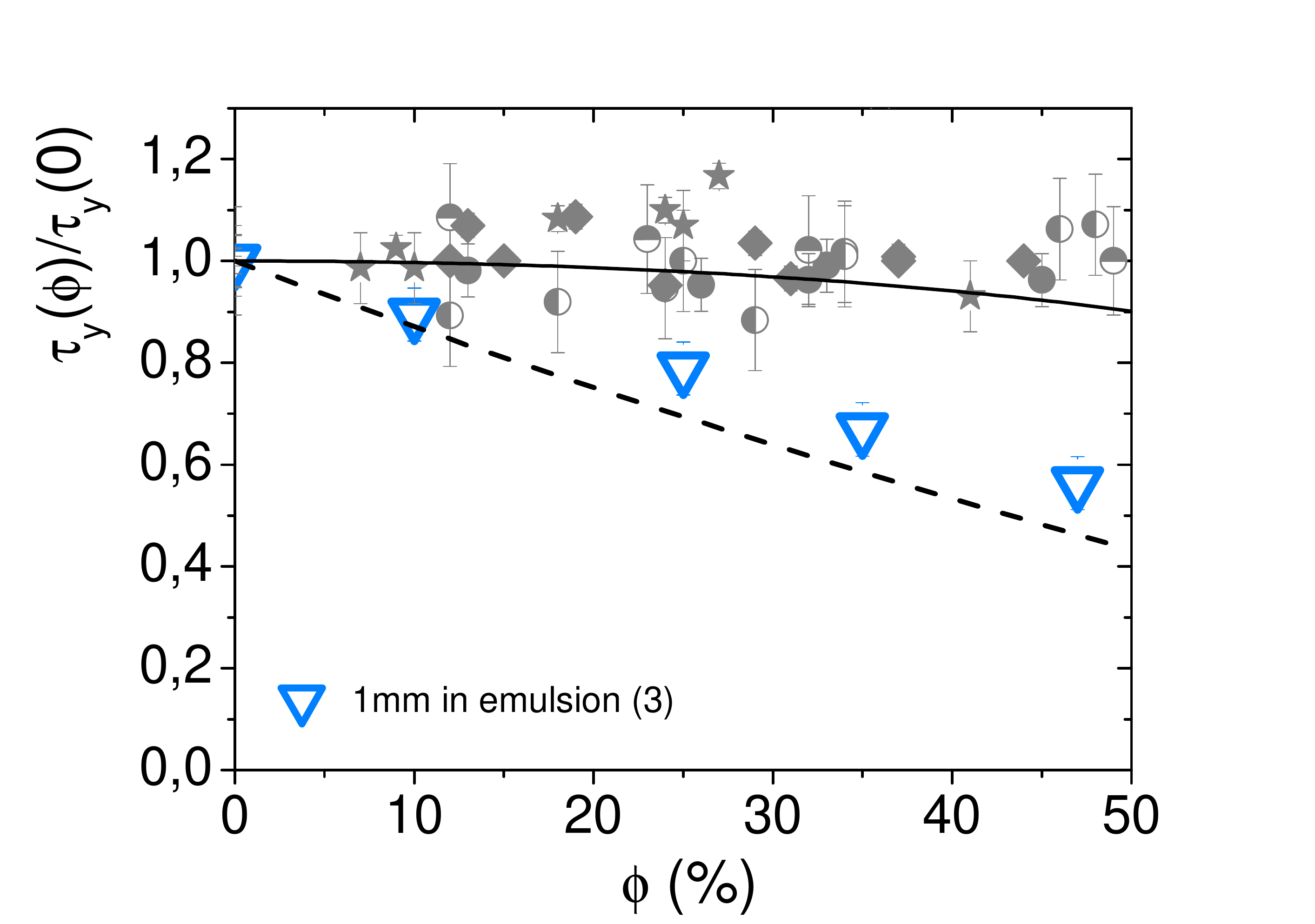}
\caption{Dimensionless yield stress as a function of the gas volume fraction, for $R=$1mm bubbles in emulsion 2. The grey symbols are all the other systems, re-plotted here for comparison. The dotted line is the theoertical limit given by equation~\ref{eq:tau_ca_inf}.\label{fig:seuils_adims+grosses_bulles}}
\end{figure}

To figure out the role played by the bubbles when flow is developed at the yield stress, we follow \cite{rust2002effects} and compare the stress exerted by the suspending emulsion on the bubbles at macroscopic yielding to the capillary stress. The capillary stress scales, as before, as the Laplace pressure. We thus compute a plastic capillary number as 
\begin{equation}
Ca_{plast}=\frac{\tau_y(0)}{2\sigma/R}
\end{equation}
For the systems presented in figure~\ref{fig:seuils_adims}, the plastic capillary number is always very small, between 0.0069 and 0.11. We can infer from this small value of $Ca_{plast}$ that the bubbles are stiff compared to the suspending emulsion at the yield stress. As a result, they remain spherical and are not significantly deformed by flow. 
\newline
To go one step further, we take advantage of the rigid behaviour of the bubbles in the suspending emulsion that allows us to use a micro-mechanical result obtained on suspensions of particles in yield stress fluids. Micro-mechanical computations~\citep{chateau2008homogenization} and dedicated experiments~\citep{mahaut2008yield} with various beads type and size in different yield stress fluids have shown that the yield stress of the suspensions can be related to their linear response $g(\phi)$ through the formula:
\begin{equation}
\frac{\tau_y(\phi)}{\tau_y(0)}=\sqrt{(1-\phi)g(\phi)}
\label{eq:Gtau}
\end{equation}
where $g(\phi)$ is the evolution with the volume fraction of any linear property of the dispersion. 
Since the bubbles are non-deformable during the yield stress measurement, they do not store any energy and the approach developed for grains holds for the bubbles. Note that the boundary condition at the surface of a particle is no-slip, whereas it is full slip at the surface of the bubble. This, however, does not modify the above result, as this boundary condition information is implicitly enclosed in the $g(\phi)$ value. In \cite{mahaut2008yield}, $G'(\phi)/G'(0)$ is chosen as a measurement of $g(\phi)$. The value of $g(\phi)$ for our rigid bubbles would need be measured on a system with a close to zero capillary number describing the linear response of the system, that is, either $Ca_{elast}\to 0$ or $Ca_{loss}\to 0$. This is not experimentally possible with our systems. We thus choose $G'(\phi)/G'(0)$ as a measurement of $g(\phi)$ and rely on the theoretical limit given by equation~\ref{eq:homog} : 
	\begin{equation}
	g(\phi)=\hat{G}_{homog}(\phi, Ca_{elast}\to 0)=\frac{5+3\phi}{5-2\phi}
	\end{equation}
The combination of this expression and equation~\ref{eq:Gtau} leads to 
\begin{equation}
\frac{\tau_y(\phi)}{\tau_y(0)}=\sqrt{\frac{(1-\phi)(5+3\phi)}{5-2\phi}}
\label{eq:tau_ca_0}
\end{equation}
which is an almost steady function which is plotted in figure~\ref{fig:seuils_adims}. It is in good agreement with the experimental data. 
\newline
To shed light on the relevancy of a plastic capillary number, we need reach higher $Ca_{plast}$ for which bubbles could be deformable at yielding of the surrounding medium. In this aim, we try to formulate suspensions with either higher yield stress or larger bubble radius. Large $Ca_{plast}$ can not be obtained by our method of mixing a foam with the emulsion: the bubbles are broken in the suspending emulsion during mixing, leading to heterogeneous suspensions and lower capillary numbers than expected (see~\cite{kogan2013mixtures} for examples). We thus prepare a series of suspensions of $R=(1\pm 0.1)\mathrm{mm}$ bubbles in emulsion 2, in which the yield stress is (40$\pm$2)Pa, thanks to a millifluidic set-up that allows us to directly inject the bubbles in the suspending emulsion. For this system, $Ca_{plast}=0.57$, and we observe that $\tau_y(\phi)/\tau_y(0)$ is a decreasing function of $\phi$, as can be seen in figure~\ref{fig:seuils_adims+grosses_bulles}. As already described for the viscosity of bubble suspensions in Newtonian fluids at high shear rates, or the elastic modulus of bubble suspensions in yield stress fluids, the relative softness of the bubbles compared to the suspending emulsion leads to a decrease of the macroscopic rheological properties.

The theoretical value of $\tau_y(\phi)/\tau_y(0)$ for fully deformable bubbles (that is, pores with no surface tension) can be computed thanks to the micro-mechanical approach leading to equation~\ref{eq:Gtau}. Although the bubbles can obviously not be seen as rigid particles, the energetic approach underlying equation~\ref{eq:Gtau} holds for fully deformable bubbles because the zero surface tension condition means that bubble deformation does not require any energy and so it does not change the global energetic balance. Substituting $g(\phi)$ by the value of $G'(\phi)/G'(0)$ for infinite capillary number (given by equation~\ref{eq:Ca_inf}) in equation~\ref{eq:Gtau}, we derive an expression for the yield stress of suspensions of fully deformable bubbles
\begin{equation}
\frac{\tau_y(\phi)}{\tau_y(0)}=\frac{(1-\phi)}{\sqrt{1+(2/3)\phi}}
\label{eq:tau_ca_inf}
\end{equation}
which is plotted in figure~\ref{fig:seuils_adims+grosses_bulles}. One can see in this figure that our experimental system is rather close to the limit case of infinitely soft (zero surface tension) bubbles in the suspending emulsion at yielding, even though the value of the plastic capillary number that we define to quantify the rigidity of the bubbles during the yield stress measurement is still lower than 1.

\section{Flow consistency}
\label{section:Flow consistency}
\subsection{Experimental results}
\begin{figure}[!h]
\includegraphics[scale=0.25]{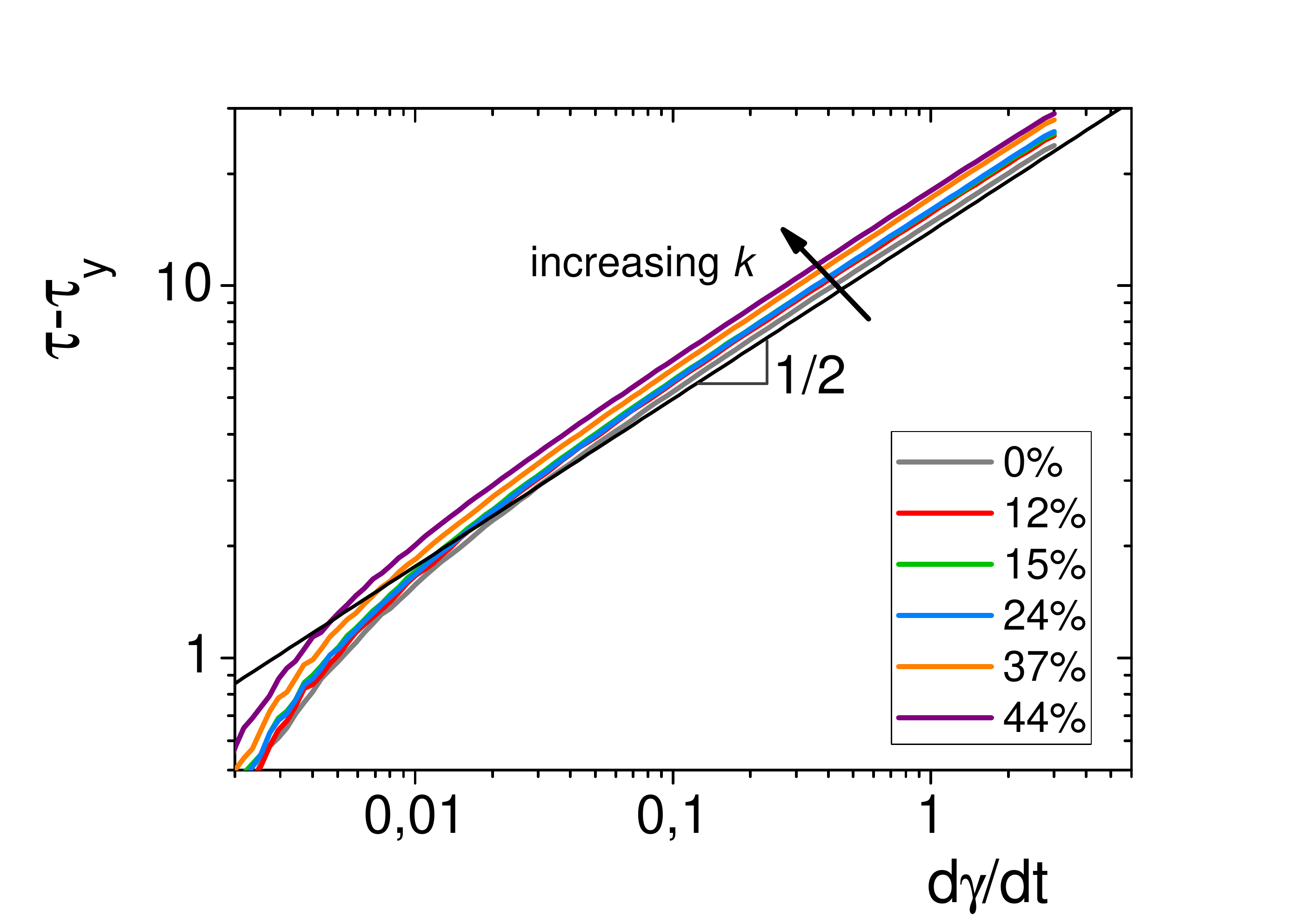}
\includegraphics[scale=0.25]{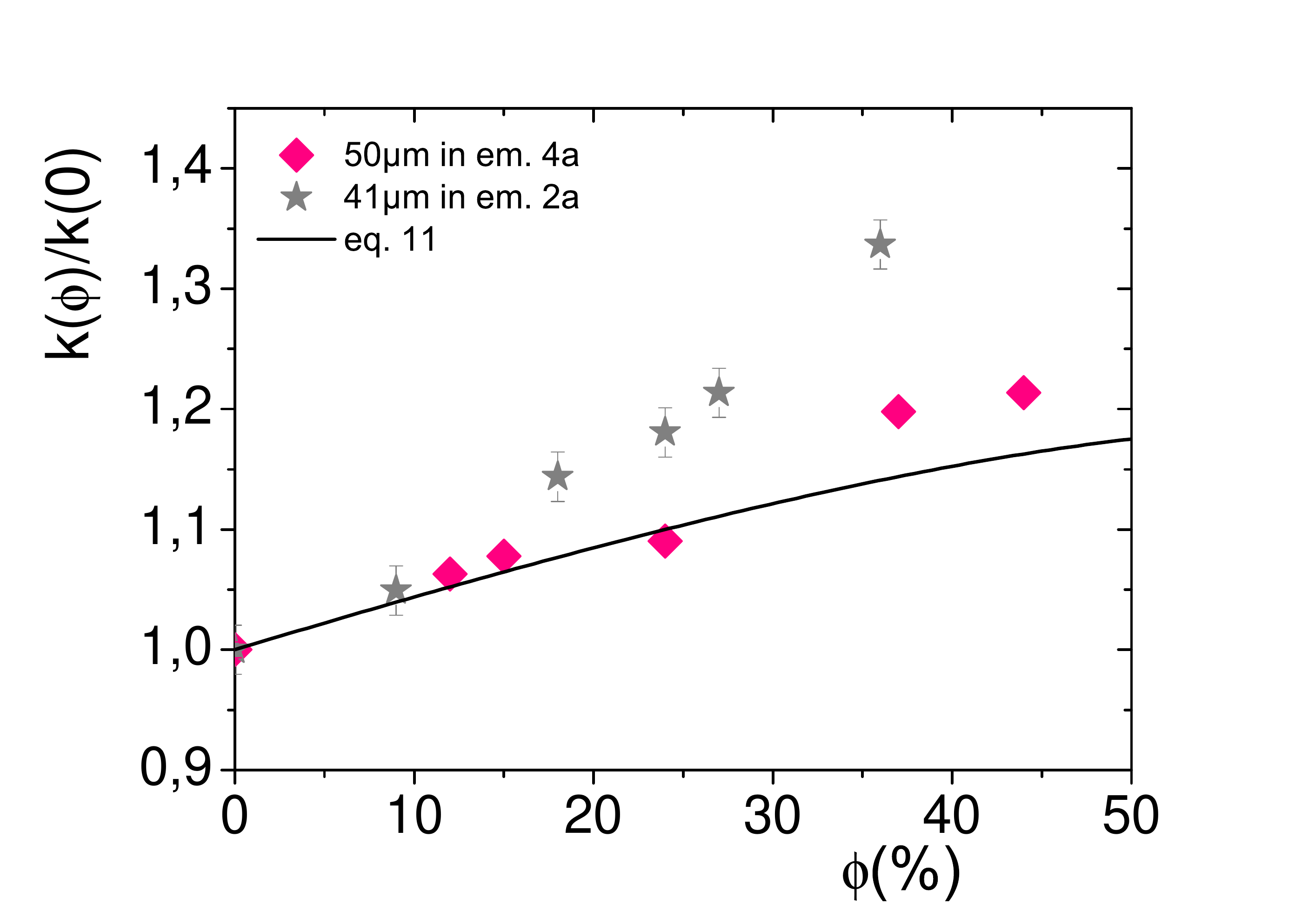}
\caption{\textit{(left)} Log-log plot of the flow curves of $R=50\mu$m bubble suspensions in emulsion 4a. The yield stress has been substracted to emphasize the common viscous power-law and the increasing consistency with $\phi$. \textit{(right)} Dimensionless consistency as a function of the gas volume fraction, for two suspensions with small bubbles. Symbols refer to the suspending emulsion composition. In both cases, $n=0.45$. The full line is a micro-mechanical estimate given by equation~\ref{eq:cons}.\label{fig:cons_adims}}
\end{figure}

The flow curve of the suspensions can be fully exploited only for those studied in parallel plates or Couette geometries, because the precise nature of flow is poorly known around a vane tool~\citep{baravian2002vane, ovarlez2011flows}. The flow curve of the suspending emulsion is well fitted to a Herschel-Bulkley law $\tau(\dot{\gamma})=\tau_y+k {\dot{\gamma}}^n$, with $n=0.45$ for our systems. We have already seen that $\tau_y$ exhibits little dependence on $\phi$. To highlight the viscous contribution to the total stress during the flow curve measurement, we plot as an example in figure~\ref{fig:cons_adims} $\tau(\dot{\gamma})-\tau_y$ as a function of $\dot{\gamma}$ for all $R=50\mu$m bubble suspensions in emulsion 4a. We can see that for a series of suspensions at various $\phi$ in a given suspending emulsion, the exponent $n$ is not modified by the presence of the bubbles. We can thus fit the flow curves with a given $n$ and extract the consistency $k(\phi)$ at each $\phi$. The dimensionless consistency $k(\phi)/k(0)$ is plotted in figure~\ref{fig:cons_adims} for two series of suspensions in different emulsions but both containing small enough bubbles to be studied in a parallel plates geometry. It is found to be an increasing function of $\phi$ for both systems.
\vfill
\subsection{Discussion}
The capillary number relevant during flow compares the total shear stress to the capillary stress scale. As can be seen on the flow curve of a suspending emulsion (figure~\ref{fig:seuil} for instance), the maximum stress applied during flow is equal or lower to three times the yield stress of the emulsion. The flow capillary number thus ranges from $0.01\leq Ca_{flow} \leq 0.3$ and is low. For this reason, we assume that the bubbles are not significantly deformed in the flow.
\newline
As we have previously done for the yield stress, because the bubbles are stiff compared to the suspending emulsion, and thus do not store any energy, we rely on a homogenization approach developed for suspensions of particles in yield stress fluids~\citep{chateau2008homogenization}. For those suspensions, the yield stress is related to the consistency by the formula:
\begin{equation}
\frac{k(\phi)}{k(0)}=\frac{{(\tau_y(\phi)/\tau_y(0))}^{n+1}}{{(1-\phi)}^{n}}
\label{eq: seuil-cons}
\end{equation}
The information about the boundary condition is in this case implicitly enclosed in the value of the yield stress of the suspensions. For our suspensions of rigid bubbles, the yield stress is well described by equation~\ref{eq:tau_ca_0}, substituting equation~\ref{eq:tau_ca_0} into equation~\ref{eq: seuil-cons} leads to
\begin{equation}
\frac{k(\phi)}{k(0)}={\left(\frac{5+3\phi}{5-2\phi}\right)}^{\frac{n+1}{2}}{(1-\phi)}^{\frac{1-n}{2}}
\label{eq:cons}
\end{equation}
This function is plotted in full line in figure~\ref{fig:cons_adims}, with $n=0.45$, which is common to both emulsions. The agreement with the experimental data is qualitatively good. Suspensions in emulsion 2a show a faster increase in the consistency than predicted, even at relatively low gas volume fraction. This may be linked to the beginning of growth in the yield stress at $\phi\sim30\%$ for this system. A reason for this difference with the other investigated systems could be the larger droplets in this emulsion (around 5$\mu$m of radius) that get trapped between bubbles even at relatively low gas volume fraction. The continuous medium hypothesis for the suspending emulsion would then become questionable.
\section{Conclusion}
We have experimentally studied the rheological properties of suspensions of bubbles in model yield stress fluids. We have seen that coupling of the suspending fluid bulk rheology to bubble deformation occurs differently depending on the stress applied to the sample. In the linear visco-elastic regime of the fluid, for the range of elastic and loss capillary numbers that we have explored, the bubbles are deformable compared to the suspending un-yielded fluid. Bubble addition then leads to a softening of the suspension. At yielding of the suspending fluid, the bubbles are generally stiff compared to the suspending fluid, in which case the yield stress of the suspension is the same as the one of the suspending emulsion. If the bubbles become softer and deformable in the suspending fluid, the yield stress of the suspensions decreases with the gas volume fraction. During flow, which could be analysed for our smallest bubble suspensions only, the bubbles are stiff in the suspending fluid and lead to an increase of the consistency with the gas volume fraction. Those distinct behaviours, entirely ruled by the gas volume fraction and bubble relative stiffness, can be quantified through different capillary numbers, relevant in each regime. Based on that approach, micro-mechanical estimates, either developed to take surface tension into account (for deformable bubbles) or adapted from developments on rigid particles (when bubbles are non- or fully deformable) prove relevant and useful. Those results are independent on the exact nature of the suspending fluid, especially on its microstructure and they should apply for any soft suspending material providing its macroscopic characteristics are taken into account to compute the relevant capillary number. The apparent limitation to soft porous materials is not a physical restriction, but it allows capillary phenomena to play a role at macroscopic length scales, whereas capillarity only needs to be considered for nanopores in standard materials.
\newline

On-going work will be dedicated to the study of high gas volume fractions, for which the emulsion can no longer be described as a continuous medium: finite size effects are expected to occur. In this regime, bubbles are pressed again one another because of geometrical constraints, and the droplets of emulsion are trapped between them. The rheology of this foamy yield stress fluid is yet to be described and understood.
\section*{Acknowledgments}
Financial support from Saint-Gobain Recherche is acknowledged. 
\section*{Appendix}
\label{appendix}
\begin{figure}[h!]
\includegraphics[scale=0.25]{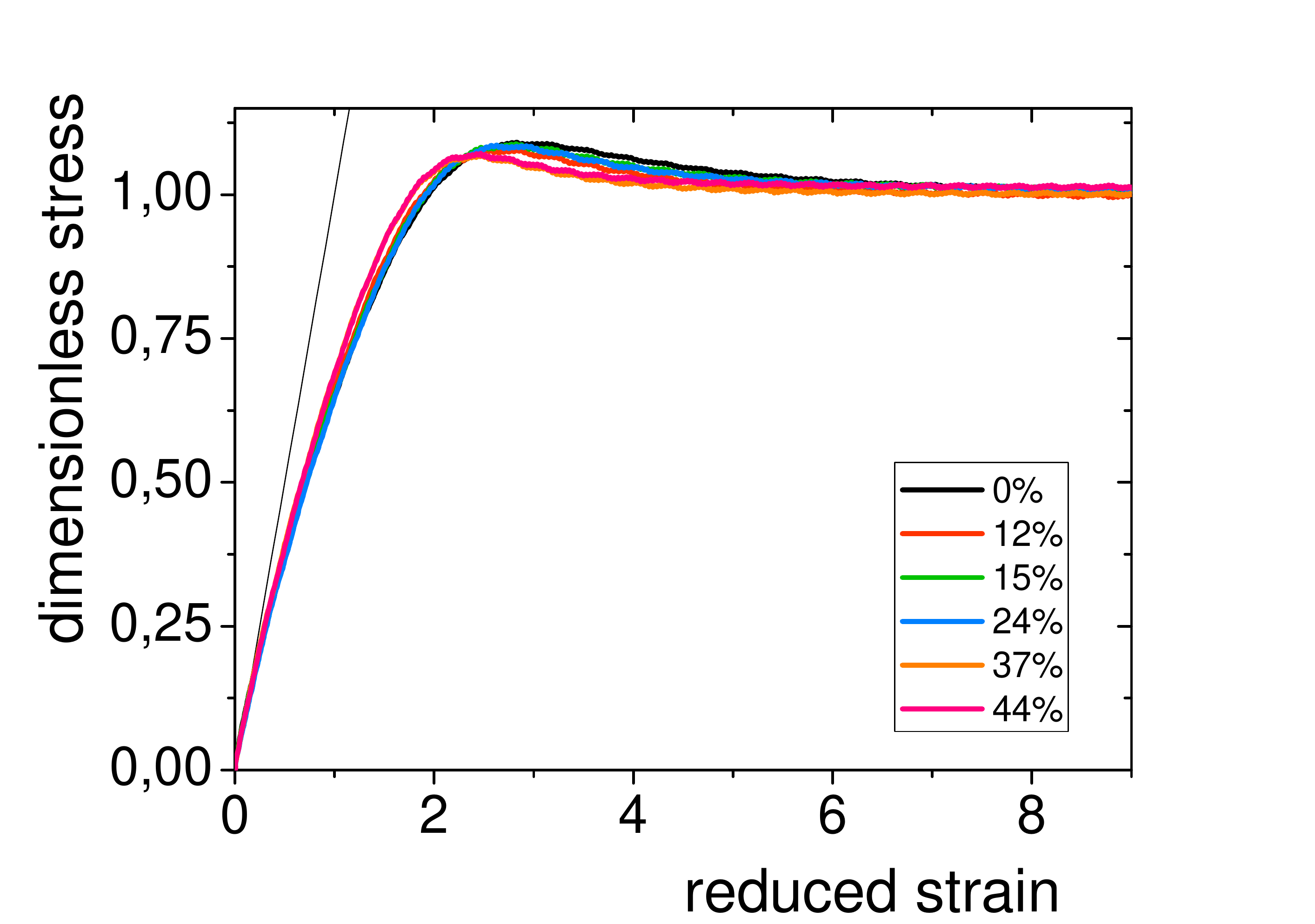}
\includegraphics[scale=0.25]{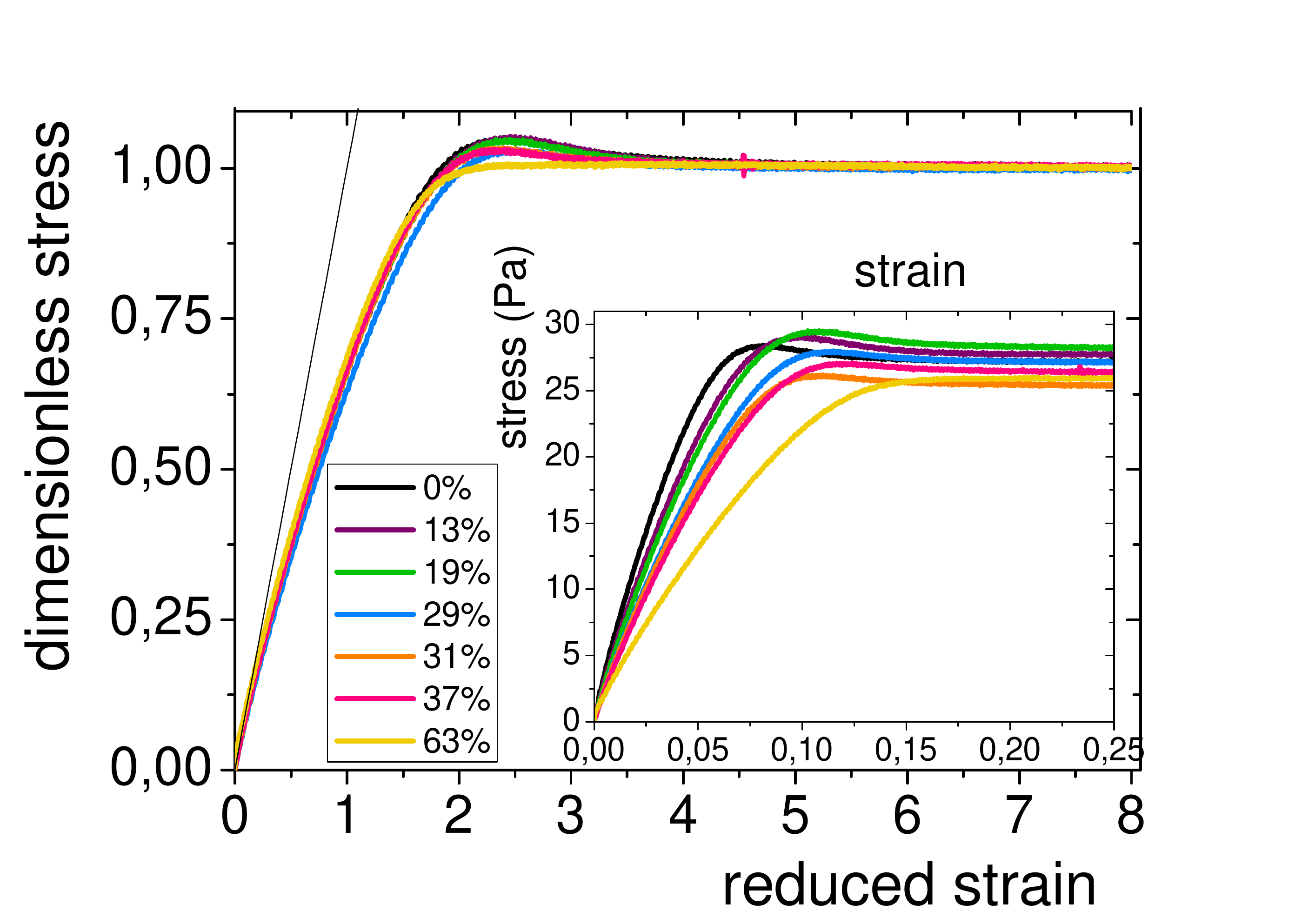}
\caption{Stress divided by the plateau yield stress, as a function of the reduced strain, for \textit{(left)} the suspensions of 50$\mu$m bubbles in emulsion 4a and \textit{(right)} the suspensions of 150$\mu$m bubbles in emulsion 4b. Inset: stress versus strain for the yield stress measurement (suspensions of 150$\mu$m bubbles in emulsion 4b), given for comparison. The decrease of $G'(\phi)$ is visible. On all graphs, the region of interest is centered on the beginning of the curves. The black line is $y=x$.\label{fig:plast}}
\end{figure}

In this appendix, we discuss in more detail the stress-strain curve obtained during the yield stress measurement. As we impose a constant shear rate to the initially at rest suspending emulsion, it first deforms elastically, and then yields, leading to a stress plateau. The very beginning of the elastic regime is linear, and the slope is equal to the elastic modulus determined during the oscillatory measurement. For larger strain, non-linear elasticity and/or creep behaviour occurs, and the stress-strain curve deviates from the initial slope. We have noted that the creep behaviour of the suspending emulsions is sensitive to the history of their flow. To make precise comparisons, we choose to present data for a system on which we have systematically made two consecutive yield stress measurements. At the end of the first measurement, the state of the material is such that their is very little creep below the yield stress, and we will thus use the second measurement for all that follows. This sample preparation should allow us to probe a possible contribution of the bubble to the softening of the suspension below the yield stress.
\newline
The suspensions of bubbles in the suspending emulsions exhibit the same shape of curves, elasto-plastic with the same plateau stress as the emulsion. However, we have seen that the elastic modulus decreases with $\phi$, meaning that the initial slope gets less and less abrupt with higher $\phi$. To isolate the possible contribution of the bubbles to the creep behaviour from that to the decreasing elastic modulus in the suspensions, we introduce a reduced strain $G'(\phi)\gamma/\tau_y(\phi)$, which re-scales the linear elastic part of all the curves. We then plot $\tau/\tau_y$ as a function of the reduced strain. In this representation, all curves have the same initial slope (equal to 1) and the same plateau stress (equal to 1, too). The result of this operation, as well as non-rescaled curves for a series of suspensions, are presented for two sets of suspensions of bubbles in figure~\ref{fig:plast}.
We observe that the beginning of all curves is well fitted by the $y=x$ function, which is consistent with the chosen scaling. We can also see that all the systems deviate from linear elasticity well before reaching yielding. However, no clear effect of the gas volume fraction is visible: for a given suspending emulsion, all the measurements seem to follow a master curve. The only impact of bubble addition on the elasto-plastic behaviour of the suspensions seems to be the decrease in $G'(\phi)$. No additional history-dependent or creep behaviour seems to be introduced by the bubbles.

\bibliographystyle{elsarticle-harv} 
\bibliography{biblio}

\end{document}